\documentstyle[preprint,aps]{revtex}
\begin{document}
\draft
\preprint{dvi file made on \today}
\title
{Angular Momentum Distribution Function \\
of the Laughlin Droplet}
\author{Sami Mitra and A. H. MacDonald}
\bigskip
%
\address
{Department~of~Physics, Indiana~University,
Bloomington~IN~47405}
%
\maketitle
\begin{abstract}
We have evaluated the angular-momentum distribution
functions for finite numbers of electrons in
Laughlin states.  For very small numbers of electrons
the angular-momentum state occupation numbers
have been evaluated exactly
while for larger numbers of electrons they have
been obtained from Monte-Carlo estimates of the
one-particle density matrix.  An exact relationship, valid
for any number of electrons,
has been derived for the ratio of the occupation numbers of the
two outermost orbitals of the Laughlin droplet and is used
to test the accuracy of the Monte-Carlo calculations.
We compare the occupation numbers near the outer edges
of the droplets with predictions based on the
chiral Luttinger liquid picture of Laughlin state edges
and discuss the surprisingly large oscillations in occupation numbers
which occur for angular momenta far from the edge.

\end{abstract}


\narrowtext

In the strong magnetic field limit all electrons in a two
dimensional electron gas occupy only the single-particle orbitals
which have the minimum quantized kinetic energy, those of
the lowest Landau level.
In the symmetric gauge, appropriate for a finite droplet
of electrons with circular symmetry,
the lowest Landau level orbitals\cite{myreview}
are labeled by angular momentum:
$\phi_m(z)=
(2\pi m!2^m)^{-1/2}~z^m~\exp(-|z|^2/4), m=0,1,2,...$
where $z_i=x_i + {\rm i}y_i$.
(We use the magnetic length,  $\ell \equiv (\hbar c/eB)^{1/2}
$, as the unit of length.)  The orbital with angular momentum
$m$ has its radial coordinate localized within $\sim \ell$ of
$R_m = \sqrt{2m} \, \ell$ while its dependence on its angular
coordinate is the same as that of an orbital with momentum
$k_m = 2 \pi m / R_m$ in a one-dimensional (1D)  electron gas
with periodic boundary conditions.  The system is therefore
strongly analogous to an interacting 1D electron gas
except that the orbitals have only one sign of momentum and
the effective `length' of the 1D gas system depends
on the momentum of the orbital.  For orbitals near the edge
of a large droplet $R_m$ is nearly fixed and the analogy with
a one dimensional gas with only one sign of momentum
(a chiral Luttinger liquid) can\cite{wenedge,myedge}
be made precise.

In the ground state of an $N$-particle non-interacting
electron gas the orbitals at energies below the
Fermi energy are occupied and all others
are empty.  For an interacting gas the single-particle state
occupation numbers fluctuate.  The change in average momentum
state occupation numbers, the momentum distribution function,  is
one of the qualitative changes due to electron-electron interactions.
In this Rapid Communication we report
on the first study of the angular-momentum distribution function
(AMDF) for strongly correlated two-dimensional electron droplets
in a strong magnetic field\cite{rezayi}.
The most compact $N$-electron droplet is a
single Slater determinant in which the orbitals with
$m=0,1,\cdots, N-1 $ are occupied; the area of this maximum
density droplet is $A_N^{MDD} \sim N 2 \pi \ell^2   \sim \pi R_N^2$.
When the electron system occupies a larger area
the average occupation numbers will be smaller than one and the ground
state of the electron system will be strongly correlated.
The best understood strongly correlated states are those
discovered by Laughlin~\cite{laughlin}:
\begin{equation}
{\psi_M}[z] = \prod^{N}_{i<j}(z_i -z_j)^M
 \exp(-\sum_k|z_k|^2/4),
\label{eq:laugh}
\end{equation}
is the ground state when the droplet
occupies~\cite{commentonconfinement,prlmdj,ajpdroplet}
an area $\sim M A_N^{MDD}$.
The M=1 Laughlin state is the maximum density droplet.
We restrict our attention here to these many-body states.

We evaluate the one-body density matrix, ${\rm n}(z,z^{\prime})$
defined by
\begin{eqnarray}
{\rm n}(z,z^{\prime})= N \int &d^2z_2&...\int d^2z_N~
\psi _M(z,z_2,...z_N)\times \nonumber \\
&\psi_M^{\ast}&(z^{\prime},z_2,...z_N) / Q_N
\label{eq:nzzpsi}
\end{eqnarray}
where $Q_N$
is the normalization integral for the $N$-electron
Laughlin wave function.
The second quantized form of Eq.~(\ref{eq:nzzpsi}) relates
${\rm n}(z,z')$ to the AMDF:
\begin{equation}
{\rm n}(z,z^{\prime})=\sum_{m=0}^{M(N-1)}
\langle n_m\rangle
\phi _m^{\ast}(z^\prime)
\phi _m(z),
\label{eq:nzzphi}
\end{equation}
To obtain Eq.~(\ref{eq:nzzphi}) we have noted that no orbital
angular momentum larger than $M(N-1)$ is ever occupied in
the $N$-electron Laughlin droplet and that the Laughlin state
is an eigenstate of total angular momentum.  ($M_{TOT} =
M N (N-1)/2$).  Eq.~(\ref{eq:nzzphi}) can be inverted to
express $\langle n_m \rangle$ in terms of ${\rm n}(z,z')$:
\begin{equation}
\langle n_m \rangle = \int d^2z \int d^2z' \phi_m^* (z)
{\rm n}(z,z') \phi_m (z').
\label{eq:amd1}
\end{equation}
For sufficiently small $N$ and $M$ the AMDF for a Laughlin
droplet can be evaluated analytically.  Some
results\cite{moreresults} for very small droplets
are listed in Table~\ref{tab:1}.

For larger droplets we have been unable to evaluate the
AMDF analytically\cite{pichard} but some properties of
the distribution function are known.  One such
property follows from the expansion of $\psi_M[z]$ in
decreasing powers of $z_1$:
\begin{eqnarray}
&\psi_M^{(N)}&(z_1,\cdots,z_N)=\nonumber \\
&[z_1&^{M(N-1)}- M(N-1) z_1^{M(N-1)-1}
\bar Z+ \cdots] \times \nonumber \\
&\exp& (- |z_1|^2 /4) \psi_M^{(N-1)}(z_2,\cdots,z_N).
\label{eq:amd2}
\end{eqnarray}
In Eq.~(\ref{eq:amd2}), $\psi_M^{(N-1)}(z_2,\cdots,z_N)$
is the $N-1$ electron droplet, $\bar Z $
is its center of mass coordinate and only the two highest powers of
$z_1$ have been retained.  Using Eq.~(\ref{eq:amd2}) in
Eq.~(\ref{eq:amd1}) we obtain
\begin{equation}
\langle n_{m_o} \rangle = N {Q_{N-1} \over Q_N} 2^{m_o} m_o!
\label{eq:amd3}
\end{equation}
where $m_o=M(N-1)$ is the angular momentum of the outermost
occupied orbital.  Similarly
\begin{equation}
\langle n_{m_o-1} \rangle = N {Q_{N-1} \over Q_N}
 2^{m_o-1} (m_o-1)! M^2 (N-1)^2 \langle {\bar Z}^2 \rangle_{N-1}.
\label{eq:amd4}
\end{equation}
It is easy to show a Laughlin droplet has definite center-of-mass
angular momentum equal to zero\cite{ajpdroplet} from which it
follows that $\langle {\bar Z}^2 \rangle_{N-1} = 2 \ell^2 / (N-1)$
and hence that for any $N$
\begin{equation}
\langle n_{m_o-1} \rangle = M \langle n_{m_o} \rangle.
\label{eq:amd5}
\end{equation}

This result can be extended to orbitals farther from the
outer edge using an argument due to Wen\cite{wenedge}.
The density, ${\rm n}(z) \equiv {\rm n}(z,z)$ far outside the
droplet may be determined up to a constant
from Eq.~(\ref{eq:nzzpsi}) and
Eq.~(\ref{eq:nzzphi}) by using a plasma analogy\cite{laughlin}.
The result for $ r= |z|  \gg R_{m_o}$ is\cite{wenedge}
\begin{equation}
{\rm n}(r) \propto \exp (-r^2/ 2 ) (r^2 / 2)^{m_o} (1 - R_{m_o}^2/r^2)^{-m}.
\label{eq:amd6}
\end{equation}
Expanding the right hand side of Eq.(~\ref{eq:amd6}) and comparing
with Eq.(~\ref{eq:amd1}) gives
\begin{equation}
\langle n_{m_o-k} \rangle = { k+M-1 \choose M-1} \langle n_{m_o} \rangle
\label{eq:amd7}
\end{equation}
which reduces to Eq.~(\ref{eq:amd5}) for $k=1$.
We remark that while Eq.~(\ref{eq:amd5}) is exact for
any number of particles Eq.~(\ref{eq:amd7}) (for $k > 1$)
becomes exact only in the limit $N^{1/2} \gg k$.  (For example,
we see that the occupation numbers listed in Table~\ref{tab:1}
do not satisfy Eq.~(\ref{eq:amd7}).  Eq.~(\ref{eq:amd7}) implies
that for $N^{1/2} \gg k \gg M $, $ \langle n_{m_o-k} \rangle
\propto k^{M-1}$.   The behavior of the AMDF in this limit is
thus consistent with Luttinger liquid\cite{haldanell}
behavior with a critical exponent related to the quantized
Hall conductance as argued on more general
grounds by Wen\cite{wenedge}.  This property of the AMDF
presumably holds
as long as the fractional quantum Hall effect occurs
and is not unique to the Laughlin state.

To examine the AMDF in the interior of the droplet and
to test how well Eq.~(\ref{eq:amd7}) is satisfied for
finite size droplets it is sufficient to evaluate ${\rm n}(z,z')$
for the case of $|z|=|z'|=r$ .
Using the explicit form of the one-particle orbitals
Eq.~(\ref{eq:nzzphi}) simplifies for this case to
\begin{equation}
{\rm n}(r,r;\theta)=\frac {1}{2\pi}
e^{-r^2/2}
\sum_{m=0}^{M(N-1)}\langle n_m\rangle
\frac {1}{m!}\frac {r^{2m}}{2}
e^{-{\rm i}m\theta}.
\label{eq:nrthetafinal}
\end{equation}
where $\theta$ is the angle between the $z$ and $z'$.
By making use of the fact that a finite number of angular momenta
have finite occupation numbers in the Laughlin droplet we
obtain an {\it exact} expression for the AMDF in terms of
a sum over a finite number of angles at a single radius:
\begin{equation}
\langle n_m\rangle =\frac {1}{f_m(r)}\sum_{j=0}^{M(N-1)}
\frac {e^{i\theta _jm}{\rm n}(r,r;\theta _j)}
{M(N-1)+1},
\label{eq:nm}
\end{equation}
where
\begin{equation}
f_m(r)=\frac {1}{2\pi m!}\frac {r^{2m}}{2}e^{-r^2/2}
\label{eq:fm}
\end{equation}
is proportional to $|\phi_m(r)|^2$ and
\begin{equation}
\theta _j=\frac {2\pi j}{M(N-1)+1}.
\label{eq:thetaj}
\end{equation}
We evaluate the AMDF numerically by combining a Monte-Carlo
evaluation of ${\rm n}(r,r;\theta_j)$ with
Eq.~(\ref{eq:nm}).

It is interesting to note that the full AMDF can
be determined, and therefore that the full
one-particle density matrix can be reconstructed,
from the angular dependence of ${\rm n}(z,z')$
at any common radius, $|z|=|z'|=r $.
This property is unique\cite{dm} to two-dimensional systems
in the strong magnetic field limit where all electrons
are restricted to a single Landau level.
However, the Monte-Carlo values of ${\rm n}(r,r;\theta_j)$
will inevitably have some statistical uncertainty.
{}From Eq.~(\ref{eq:nm}) we see that the resulting
uncertainty in the occupation number will be a minimum
when $r$ is near the maximum in $f_m(r)$ which occurs at
at $r=\sqrt {2m}= R_m$.  Typical uncertainties in
$\langle n_m \rangle$ become very large unless $r$ is near
$R_m$.  For all the numerical results reported
below we estimated $\langle n_m\rangle$ values from the
the angular dependence of the density matrix at $r$ near $R_m$.

The Monte-Carlo calculation evaluates the complex
function ${\rm n}(r,r;\theta)$ by a
Metropolis~\cite{metropolis} sampling
of the positions of particles $2$ through $N$.
The weighting factor is
\begin{equation}
W(z_2,...,z_N)=\, \prod_{1<l<m}^{N}
|z_l -z_m|^{2M}
 \exp(-\sum_{k>1}|z_k|^2/2).
\label{eq:weight}
\end{equation}
With this weighting factor
\begin{eqnarray}
{\rm n}(r,r;\theta) = N{Q_{N-1} \over Q_N}
\langle e^{-r^2/2}\prod_{j>1}^{N}
\lbrace \! &(&\! re^{-i\theta}\! -\!
r_je^{-i\theta _j})\nonumber \\
&\times&
(r\! -\! r_je^{i\theta _j})\rbrace^M\rangle.
\label{eq:nraveraged}
\end{eqnarray}
Thus $n(r,r;\theta)$ is determined up to a constant
which is independent of both $\theta$ and $r$.  We
determine $Q_{N-1}/Q_N$ by requiring that the
integral of the diagonal elements of the density-matrix
over position equal $N$
($\int d\vec r\> {\rm n}(r,r;\theta=0)= N)$.
${\rm n}(r,r;\theta _j)$ was evaluated in this way
at a set of radii separated by $0.5 \ell$.
$\langle n_m \rangle$ was estimated by averaging
results over different values of $r$
weighted by the factor $e^{-(r-R_m)^2/2}$.

Results for the occupation numbers obtained in this
way are illustrated in Fig.~(\ref{fig:fig1}) for
$M=3$, $M=5$, and $M=7$ Laughlin droplets with
$N=25$, $N=20$ and $N=15$ respectively.
Monte Carlo calculations were also carried out for smaller
size droplets where comparisons with
analytic results could be
made.  The two main features apparent in these results
are the large oscillations in the occupation numbers
in the interior of the droplets and
the rapid decline in the occupation numbers as the outermost
orbitals are approached.  For the three cases illustrated
$m_o=72, 95$ and $98$ and the occupation numbers
for the outermost droplets are $1.07\times 10^{-2}$,
$2.3\times 10^{-4}$ and $2.6\times 10^{-5}$
respectively.  In Fig.~(\ref{fig:fig2})
we compare the angular momentum state occupation numbers
near the edge of the droplet with Eq.~(\ref{eq:amd7}), which
we find to be accurately satisfied for the droplet sizes
studied.  We conclude that Luttinger-liquid-like behavior should
be visible\cite{kane} in the low-energy properties
even for quite small Laughlin droplets.
For example for the $M=7$, $N=15$, Laughlin droplet
$m_o=98$, and the Monte Carlo calculation gives
$\langle n_{93} \rangle/ \langle n_{98} \rangle
=  423 \pm 15$ compared to the value $462$ implied by
Eq.~(\ref{eq:amd7}).  For the droplet sizes illustrated
Eq.~(\ref{eq:amd7}) begins to fail badly for $k$ larger than
$\sim 10$.  Away from the edge the occupation numbers reach
a maximum value and oscillate as the
interior is approached.  In the limit of extremely large
droplets it is known from the plasma analogy for Laughlin
wave functions that the density deep in the interior
approaches $(2 \pi M)^{-1}$, from which it follows that
$\langle n_m \rangle $ approaches $M^{-1}$.
Our Monte-Carlo calculations show that this limit is
approached slowly as $N$ increases.  The largest occupation
numbers occur near the edge of the droplet and the
relative excess in this region is larger for larger $M$.
For the droplets illustrated
the largest occupation numbers are $\langle n_{50} \rangle=
0.52 \pm 0.02$ for $M=3$,
$\langle n_{72}\rangle=0.47 \pm 0.04$ for $M=5$
and $\langle n_{75}\rangle=0.43 \pm 0.03$ for $M=7$.
The oscillations in occupation number are related to
the oscillations in charge density expected near the
edge of a finite 2D plasma\cite{ferrari} but are much more
pronounced since the density averages occupation
numbers of orbitals with $R_m$ near $r$.

As we see from the plots of $\langle n_m\rangle$,
the statistical uncertainties, obtained by averaging
over independent runs, are smaller closer to the edge of the
droplet. We believe that this is because fewer orbitals
contribute importantly to the one-particle density matrix.
In fact the occupation numbers
for orbitals very close to the edge are actually
more reliably calculated by evaluating the angular dependence
of the density-matrix at $r$ substantially larger
than $R_{m_o}$.  In closing we remark that it is in principle
possible to determine the occupation numbers
uniquely if the radial dependence of the particle-density
is known precisely.  However, this procedure is extremely
ill-conditioned and we believe that it is practical only
where the particle-density is known analytically.  We have
found it to be impossible to determine the AMDF accurately from
Monte-Carlo particle densities even for quite small droplets.

This work was supported by the National Science Foundation
under grant DMR-9113911.  The authors are grateful to
Mats Wallin for generous and helpful advice on the
Monte Carlo calculations. AHM acknowledges informative
conversations with X.-G. Wen, and E.H. Rezayi.

\newpage

\begin{figure}
\caption{
Occupation numbers for $M=3$, $M=5$, and $M=7$,
Laughlin droplets with $N=25$, $N=20$, and $N=15$
respectively plotted vs. $m= R_m^2/2$.
The solid line is the particle density
in $(2 \pi \ell^2)^{-1}$ units plotted vs.$r^2/2$
calculated from the occupation
 numbers using Eq.~(3).
The dashed line shows the particle density calculated
directly; the difference in these two quantities is a
measure of the error introduced in extracting the occupation
numbers from the density matrix.
\label{fig:fig1}}
\end{figure}

\begin{figure}
\caption{
Comparison of Monte Carlo
occupation numbers near the
edge with the formula derived by Wen, Eq.~(10).
\label{fig:fig2}}
\end{figure}

\begin{table}
\caption{Angular momentum state occupation numbers for
some representative small $N$ Laughlin droplets}
\begin{tabular}{cccccccc}
$M,N$&$m=0$&$m=1$&$m=2$&$m=3$&$m=4$&$m=5$&$m=6$\\
\tableline
3,2&1/4&3/4&3/4&1/4&0&0&0\\
3,3&7/31&9/31&18/31&22/31&21/31&12/31&4/31\\
5,2&1/16&5/16&10/16&10/16&5/16&1/16&0\\
\end{tabular}
\label{tab:1}
\end{table}

\begin{references}
\bibitem{commentonconfinement} The area occupied by a droplet
depends on the magnetic field strength and on the nature of
the potential which confines the electrons to a finite area.
For typical confinement potential and a finite number of
electrons there will be a finite range of magnetic fields
over which each of the Laughlin droplet states becomes
the ground state.
\end{references}
\end{document}